\documentstyle[aps,epsf]{revtex}
\begin{document}
\author{Anne-Marie Dar\'{e}\cite{dare0}, and Gilbert Albinet}
\title{Magnetic properties of the three-dimensional Hubbard model at half
filling.} 
\address{I.R.P.H.E., UMR 6594 \\
Universit\'{e} de Provence, 
Campus de St J\'erome, Case 252 \\
13397 Marseille Cedex 20, France\\}
\date{14 septembre 1999}
\maketitle

\begin{abstract}
We study the magnetic properties of the 3d Hubbard model at half-filling in the
TPSC formalism, previously developed for the 2d model. We focus on the N\'eel
transition approached from the disordered side and on the paramagnetic phase. 
We find a very good quantitative agreement with Dynamical Mean-Field 
results for the isotropic 3d model.
Calculations on finite size lattices also provide satisfactory comparisons with
Monte Carlo results up to the intermediate coupling regime. 
We point out a qualitative difference between the isotropic 3d case, and the
2d or anisotropic 3d cases for the double occupation factor. Even for this
local correlation function, 2d or anisotropic 3d cases are out of reach of DMF: this 
comes from the inability of DMF 
to account for antiferromagnetic fluctuations, which are
crucial. 

\end{abstract}


\section{Introduction}

The problem of strongly correlated fermions is a long-standing opened question.
Recently progress in the resolution of the simplest model, namely the
Hubbard model in $d=\infty$ by the Dynamical Mean Field approach
(DMF),\cite{Georges3} \cite{Georges2} marked a real breakthrough in the field.
Nevertheless, the lack of spatial dependence and the difficulties in including
the $1 /d$ corrections,\cite{Georges2} and the low dimensionality or the
anisotropic character of the experimental systems, justify the ongoing
research on more realistic dimensions $d=2$ and 3. 
In low dimension, especially $d=2$, one expects a strong ${\bf k}$-dependence
of the self-energy that is absent from $d=\infty$ calculations. In this paper
we study the 3d isotropic Hubbard model, using a formalism previously
developed in 2d,\cite{Vilk}\cite{Vilk2} named TPSC for Two Particle Self
Consistent. In 2d, the TPSC approach has been shown to give reliable results
in comparisons with  Monte Carlo calculations for one and two-particle
properties, static and dynamical quantities, for a whole range of parameters:
for a general band filling, on the whole temperature range accessible to Monte
Carlo simulations (but probably not at very low temperature at half-filling
in $2d$), from low up to intermediate value of $U/W$, for
the nearest neighbor model as well as for the $Utt'$ case.\cite{Veilleux} The
strengths of this approach rely on the respect of several benchmark
properties: it fulfills the Pauli principle and is conserving in the Baym and
Kadanoff sense.\cite{Baym} It was shown in 2d to satisfy the Mermin-Wagner
theorem. Furthermore the TPSC approach is appealing due to its simplicity: the
vertices retained in the particle-hole channels are static and local.

Apart from the 2d case, 
the highly anisotropic 3d case (inter-plane hopping much less than in-plane
one) has been studied previously.\cite{dare} It was found that the
antiferromagnetic order can be established in the regime of incoherent 3d
one-particle motion and the crossover between 2 and 3d was studied in details.
In this paper we want to explore the magnetic properties of the isotropic 3d
model, focusing on the paramagnetic phase and on the N\'eel
transition approached from the disordered side.
We make comparisons with other approaches, including Monte Carlo
results, and DMF approach for the hypercubic lattice. We also stress the
qualitative difference between isotropic and anisotropic 3d cases, even
for a local function as the double occupation factor.  

The paper is organized as follows: In Sec. II we review the TPSC formalism
for the particle-hole properties, namely spin and charge susceptibilities. 
Sec. III concerns the N\'eel transition, and Sec. IV is
devoted to the temperature dependence of the double occupation factor from
which all the magnetic properties follow. Finally we summarize our conclusions
in Sec. V. An appendix details the calculation of the double occupation factor
for a simple two-site model that enlightens the high-temperature
lattice results.

\section{Two-Particle Self-Consistent formalism }

We start from the Hubbard Hamiltonian,

\begin{equation}
H=-\sum_{<ij>\sigma }t_{i,j}\left( c_{i\sigma }^{\dagger }c_{j\sigma
}+c_{j\sigma }^{\dagger }c_{i\sigma }\right) +U\sum_in_{i\uparrow
}n_{i\downarrow \,\,\,\,} -\mu\sum_{i \sigma}n_{i\sigma\,\,\,\,} , 
\label{Hubbard} \end{equation}
where the operator $c_{i\sigma }$ ($c_{i\sigma }^{\dagger }$) destroys
(creates) an electron of spin  $\sigma $ at site $i$, $n_{i\sigma}$ is the
density operator, and $t_{i,j}$ is the  symmetric hopping matrix. The on-site
repulsion $U$ is the screened Coulomb interaction, $\mu$ the chemical
potential.  For a while hopping parameters will be kept general. Indeed the
Two-Particle Self-Consistent formalism (TPSC) approach only requires a local
interaction but can be applied to a general band structure. 

Let us sketch the TPSC approach, fully
developed in Ref. \cite{Vilk} and \cite{Vilk2}. The TPSC can be 
formulated in a closed form in the functional integral formalism.\cite{Vilk2}  Here we
shall concentrate on the two-particle properties, and obtain the
useful relations in the equation of motion approach.
In the presence of a time, space and spin varying
external field $\phi$,  a generalized Green function can be
defined:\cite{Kadanoff} \begin{equation}
G(r_i,r_j,\tau_i;\phi)={ Tr \left[ e^{-\beta H} T_\tau S(\beta) 
c_{i\sigma }^{\dagger } (\tau_i)  c_{j\sigma }(\tau_i)
\right]  \over 
Tr \left[ e^{-\beta H} T_\tau S(\beta)   \right]
} ,
\label {greenfun} 
\end{equation}
where $T_\tau$ means imaginary time ordering and
\begin{equation}
\ln S(\beta) = - {\int_0 ^\beta}   d \tau   \sum_{l,\sigma} 
c_{l\sigma }^{\dagger } (\tau)c_{l\sigma }(\tau) 
\phi_\sigma (l,\tau) \,\,\,\, . 
\end{equation}
All the field dependence is in $S(\beta)$ and
\begin{equation}
c_{l\sigma }^{\dagger } (\tau)= e^{\tau H } c_{l\sigma }^{\dagger } 
e^{-\tau H \,\,\,\,}.
\end{equation}
To obtain the spin and charge susceptibilities, the equation of motion for 
$ G$ has to be evaluated. Then in the 
linear response theory, the functional derivative with respect to $\phi$
is taken at zero field leading to:
\begin{equation}
\delta  \left( \frac {\partial \left< c_{i\sigma }^{\dagger } c_{j\sigma
}\right>} {\partial \tau_i} \right )=\left( \delta \phi_\sigma (i,\tau_i)
-\delta \phi_\sigma (j,\tau_i) \right) \left. \left< c_{i\sigma }^{\dagger }
c_{j\sigma}\right> \right|_{\phi=0}
+U \delta \left< c_{i\sigma }^{\dagger } c_{j\sigma} n_{l \bar{\sigma}}\left(
\delta_{i,l} - \delta_{j,l}\right) \right> + \delta C \,\,\,\,  ,
\label{derivee}
\end{equation}
with the kinetic part
\begin{equation}
\delta C = \frac 1 {N} \sum_{\bf kq} \left(\epsilon_{\bf k} -\epsilon_{\bf k+q}
\right) e^{-i{\bf k}\bf r_i}  e^{i ({\bf k+q})\bf r_j} \delta \left<
c_{{\bf k}\sigma }^{\dagger } c_{{\bf k+q}\sigma} \right>\,\,\,\, .
\label{deltac} \end{equation}
All the operators in Eqs. (\ref{derivee}) and (\ref{deltac}) are at the same
time $\tau_i$ and  $\left < ...\right>$ is a short-hand notation for averaging
with  $S(\beta)$ as in Eq. (\ref{greenfun}). $\bar{\sigma}$ is opposite
to $\sigma$, ${\bf k}$ is a wave vector of an $N$-site lattice, and
$\epsilon_k$ is the Fourier transform of $t_{i,j}$. 

The ansatz of the TPSC approach is to write, for the term proportional to
$U$ in Eq.(\ref{derivee}), that
the following  approximation retains the main correlation effects:
\begin{equation} \delta \left< c_{i\sigma }^{\dagger } c_{j\sigma}  n_{l
\bar{\sigma}}  \right> \simeq \delta \left( g_{\sigma \bar{\sigma}}(l,\tau_i) 
\left< c_{i\sigma }^{\dagger } c_{j\sigma} \right> 
\left< n_{l \bar{\sigma} } \right> 
 \right)\,\,\,\, \,\,\,\,\,\,\,\, {\rm for }\,\,\,\, l=i \,\,\,\,{\rm or}\,\,\,\,l=j
\,\,\,\,  , \label{ansatz}
\end{equation}
with the following definition:
\begin{equation}
g_{\sigma \bar{\sigma}}(l,\tau_i)=\frac {\left< n_{l\sigma }(\tau_i)
n_{l \bar{\sigma}}(\tau_i)\right>}{\left< n_{l\sigma }(\tau_i)\right>
\left<n_{l \bar{\sigma}}(\tau_i)\right>}
\label{gupdn}
\end{equation}
Eq. (\ref{ansatz}) is exact when $i=j$. The expansion of the functional
derivative with respect to $\phi$ leads to
\begin{equation} 
U \delta \left< c_{i\sigma }^{\dagger } c_{j\sigma} \left( 
n_{i \bar{\sigma}} - n_{j \bar{\sigma}}\right) \right> = 
U  \left. \left< c_{i\sigma }^{\dagger } c_{j\sigma} \right>
\right|_{\phi=0} \left\{\left. \left< n_{i\bar{\sigma} } \right>
\right|_{\phi=0} \left( \delta g_{\sigma \bar{\sigma}} (i,\tau_i) - \delta
g_{\sigma \bar{\sigma}} (j,\tau_i)\right)+
\left. g_{\sigma \bar{\sigma}}\right|_{\phi=0}
\left( \delta\left< n_{i\bar{\sigma} } \right> -\delta\left< n_{j\bar{\sigma}
} \right>\right) \right\}\,  ,
\label{truc}
\end{equation}
since at zero field in the paramagnetic phase, the coefficient of the term
proportional to $\delta\left< c_{i\sigma }^{\dagger } c_{j\sigma} \right>$
cancels.

To get the spin susceptibility it remains to take
$ \phi_\sigma (i,\tau_i)=\sigma \phi (i,\tau_i)$, then to
time- and site-Fourier transform Eq.(\ref{derivee}), using 
Eqs.(\ref{deltac}) to (\ref{truc}).
From the definition  
\begin{equation}
\chi_{sp}(q,i\omega_q) = - \frac 1 N \sum_{k} \frac {\delta \left< 
\sum_\sigma \sigma c_{k \sigma}^\dagger c_{k+q \sigma} \right>_{i
\omega_q}}{\delta \phi(q,i\omega_q)}  
\end{equation}
the preceding equations lead to an effective RPA expression for the spin
susceptibility 
\begin{equation}
\chi_{sp}(q,i\omega_q) = \frac {\chi_0 (q,i\omega_q)}{1- \frac
{1}{2} U_{sp}\chi_0 (q,i\omega_q)}\,\,\,\,  , 
\label{suscepsp}
\end{equation}
with 
\begin{equation}
U_{sp}=
g_{\uparrow\downarrow} U\,\,\,\,  . 
\label{uspin}
\end{equation}
$g_{\uparrow\downarrow}$ is the time- and site-independent version of
Eq.(\ref{gupdn}).
This $q$ and $i\omega_q$ independence follows from the cancellation of the term
proportional to  $\delta g_{\sigma\bar{\sigma}}$ which does not depend on the
spin index, and disappears by summation on $\sigma$: $\sum_\sigma \sigma \delta
g_{\sigma\bar{\sigma}} =0$.
As will be recalled later, $g_{\uparrow\downarrow}$ can be evaluated in a
self-consistent equation dictated by sum rules.
For the charge susceptibilities the term proportional to $\delta
g_{\sigma\bar{\sigma}}$ doesn't cancel, and one obtains a 
${\bf q}$ and $i\omega_q$-dependent renormalized interaction. However further
simplification consisting in taking a ${\bf q}$ and $i\omega_q$-constant
renormalized vertex for charge $U_{ch} =(g_{\uparrow\downarrow}+\delta
g)U$, where $\delta g$ can be self-consistently evaluated using exact sum
rules, has been shown to give reasonable results in comparison to Monte Carlo
simulations. The sum rules to determine self-consistently $U_{sp}$ and
$U_{ch}$ follow from the the Pauli principle and the fluctuation-dissipation
theorem, here expressed in Matsubara formalism 
\begin{equation} 
\left< S_i^zS_i^z \right> = \left< (n_ \uparrow - n_\downarrow)^2 \right> = n
-2\left<n_\uparrow n_\downarrow\right>
=\frac 1{\beta N}\sum_{{\bf q}, {i\omega_q}}
\frac{\chi _0({\bf q}, {i\omega_q})}{1-\frac 12U_{sp}\chi _0(%
{\bf q}, {i\omega_q})},  \label{sumSpin}
\end{equation}
and
\begin{equation}
\left< \delta n_i\delta n_i\right> =
n+2\langle n_{\uparrow }n_{\downarrow }\rangle -n^2=\frac 1{\beta N}\sum_{%
{\bf q}, {i\omega_q}}\frac{\chi _0({\bf q}, {i\omega_q}))}{1+\frac 12U_{ch}\chi _0(%
{\bf q}, {i\omega_q})},  \label{sumCharge}
\end{equation}
where $\beta \equiv 1/T$, $n=\langle n_{\uparrow }\rangle +\langle
n_{\downarrow }\rangle $, and
$\delta n_i= \sum_\sigma n_{i \sigma}-n$. The necessity to have two different
irreducible vertices for spin and charge is imposed by the Pauli principle.
Simple RPA for which the spin and charge susceptibilities differ only
by a sign in the denominator cannot fulfill these previous rules.

Although the
susceptibilities have an RPA functional form, the physical properties of the
theory are very different from RPA because of the self-consistency
conditions for determining $U_{sp}$ and $U_{ch}$.
This approach has been shown\cite{Vilk} to take into account both the local
quantum renormalization effects (Kanamori-Brueckner screening) which reduce 
the Coulomb repulsion, and long-wavelength
thermal fluctuations, giving a phase transition only at
zero-temperature in two dimensions. 
This theory is self-consistent at the two-particle level (hence its
name), in contrast with FLEX\cite{FLEX} for example, which is
self-consistent at the one-particle level. 

In this paper we shall concentrate on magnetic properties
for the nearest-neighbor model on a 3d-simple cubic lattice with a spacing
$a=1$, for which the dispersion relation reads
\begin{equation}
\epsilon _{{\bf k}}=-2t\sum_{i=1}^3 \cos k_i ,
\label{reldisp} 
\end{equation}
with a bandwidth $W=12t$.
The numerics only concern the half-filled band case, hence $\chi _0({\bf
q}, 0)$ is maximum at ${\bf Q} =(\pi,\pi,\pi)$ due to the perfect nesting. 
The calculations are done on a variable
mesh for the Brillouin zone: $16^3$ on the coarse one,  $16^3$ more points near
${\bf Q}$, but as will be detailed below, some integrals are analytically
evaluated, and the non-interacting susceptibility is finely evaluated at low
temperature. We also present calculations on finite-size lattice for
comparisons with Monte Carlo results. Hereafter $t$ will be the energy unit,
and we set $k_B=1$.

\section{The antiferromagnetic transition}

The TPSC formalism has been shown to
give reliable results far from the transition and also in the part of
the critical regime that is accessible to Monte Carlo calculations. Important
constraints such as the Mermin-Wagner theorem are fulfilled by the TPSC
approach. Indeed in 2d as the temperature goes down, $U_{sp}$ decreases, thus
preventing a finite-temperature magnetic transition. However deep in the low
temperature regime at half-filling in 2d, the relation $U_{sp}= U g_{\uparrow
\downarrow}$ breaks down: there is no reason to believe that there is no
double-occupied site in the ground state.\cite{Vilk2} In the highly
anisotropic three-dimensional case with a small inter-plane hopping $t_z \ll t$,
a finite-temperature transition can be stabilized. Here we shall concentrate on
the isotropic case and make comparisons with other approaches developed for
the 3d or for the infinite-dimensional Hubbard model.

In 3d, at half filling, the
self-consistent Eq.(\ref{sumSpin}) from which the magnetic
properties will be evaluated, can be rewritten  
\begin{equation}
1-\frac 1 2 g_{\uparrow\downarrow}
=T \int \frac {d^3q}{(2\pi)^3} \chi_{sp}^{as}({\bf q},0) +
T \int \frac {d^3q}{(2\pi)^3}\left(
\chi_{sp}({\bf q},0)-\chi_{sp}^{as}({\bf q},0)\right)+T \int \frac
{d^3q}{(2\pi)^3} \sum_{i\omega_q\neq 0} \chi_{sp}({\bf q},i\omega_q) , 
\label{autoco} 
\end{equation}
where $\chi_{sp}^{as}({\bf q},0)$ is the asymptotic form of $\chi_{sp}({\bf
q},0)$ obtained from a Taylor expansion around ${\bf Q}$ of the
denominator and defined on a domain of radius $\Lambda$ around ${\bf Q}$. This
enables an analytic evaluation of the diverging part of the susceptibility.
Defining $U_{mfc}=2 / \chi_0({\bf Q},0)$ the mean-field critical value of $U$
at which the transition occurs in a RPA treatment, and $\delta U=
U_{mfc}-U_{sp}$ the small energy scale that measures the proximity to the
phase transition in the TPSC approach, the asymptotic form of the spin
susceptibility simply reads \begin{equation}
\chi_{sp}^{as}({\bf q},0)= \frac 2{\delta U} \frac {1}{1+ \xi^2 \left[
\left(q_x-\pi \right)^2 +\left(q_y-\pi \right)^2+\left(q_z-\pi
\right)^2\right]}  \,\,\,\,\,\,  , 
\label{chiasymp}
\end{equation}
where 
$\xi$ is the magnetic correlation length defined in terms of a
microscopic length $\xi_0$:
\begin{equation}
\xi^2=\xi_0^2 \frac {U_{sp}}{\delta U} \,\,\,\, 
\label{ksi}\end{equation}
\begin{equation}
\xi_0^2= \frac 1 4 U_{mfc} \left| \frac {\partial^2 \chi_0}{\partial q_x^2}
\right|_{{\bf q}={\bf Q}} \,\,\,\,\,\,   .
\label{ksi02}
\end{equation}
The N\'eel transition is reached from above when $U_{sp}\rightarrow
U_{mfc}^-$.   
The TPSC approach validity is restricted from low up to intermediate value of 
the ratio $U/W$. Indeed it has been
shown, and will be confirmed below, that $U_{sp}$ saturates at strong
coupling while one would expect $U_{sp}$ to have a maximum as a function of
$U$, as suggested by RPA at low repulsion and by the Heisenberg model limit at
strong repulsion. Nevertheless the reliability for $U \lesssim W$ in 2d
has been established by comparisons with Monte Carlo calculations, and gives
some confidence in 3d for the intermediate coupling regime. 

Fig.1 presents $T_N$ as a function of $U$ in units of 
the hopping parameter $t=1$.
TPSC results are presented along with Monte Carlo,\cite{Scalettar} 
RPA, dynamical mean field results on an
hypercubic lattice,\cite{Jarrell1}\cite{Jarrell2} spin-fluctuation 
formalism,\cite{Singh} and spin Hamiltonian 
approach.\cite{Szczech}\cite{Tusch} The last one is a mapping of the Hubbard
model onto an Heisenberg model with arbitrary-range, U-dependent exchange
integrals $J_{ij}(U)$, treated in an Onsager reaction field method.
The DMF results are scaled for comparisons with $3d$ by $\sqrt{< \epsilon_k^2
>}$ (details will be given in next section). 

Usual RPA predicts an exponential behavior:
\begin{equation} 
T_N^{RPA}= \Gamma \exp \left(-\frac 1{N(0)U}\right)\,\,\,\,  ,
\end{equation}
where $\Gamma =3.8 \pm 0.1$, and $N(0) \simeq 0.14$ is the 3d DOS at the
Fermi level. Due to the same functional form, the
TPSC formalism also predicts an exponential form for $T_N$ with $U$
replaced by $U_{sp}$.
The saturation for $U \gtrsim W$ of $U_{sp}$ to the value $\simeq 3.5$ 
leads to the saturation of $T_N$ as can be seen in
Fig.1. $U_{sp}=U g_{\uparrow\downarrow}$ is not the correct ansatz at
large $U$.

Recently there have been a few controversies concerning the N\'eel temperature.
One of these concerns the relevance of the RPA at weak
coupling.\cite{vanDongen}\cite{Tahvildar} 
A discrepancy was found in an infinite-dimensional approach with $1/d$
corrections in Ref.\cite{vanDongen}, and in pure 3d formalism in
Ref.\cite{Tahvildar}. From Fig.1 it appears that the TPSC curve merges with
RPA results, but only at very small coupling. The modified
Stoner criterion used to obtain the N\'eel temperature in Ref.\cite{Tahvildar}
uses a maximally crossed diagram series which was shown to be not reliable
close to half-filling in 2d.\cite{Chen} Up to $U \simeq W$, TPSC and
DMF approaches are very close.

The second controversy \cite{Szczech} concerns the interpretation of the Monte
Carlo results: Hasegawa\cite{Hasegawa} has argued that the N\'eel temperature
extracted from the Monte Carlo simulations was overestimated, in fact it would
be rather a mean-field temperature than a true $T_N$. A indication of this can
be seen at large $U$, where the Monte Carlo is closer to the mean-field result 
$T_N^{MF} \simeq 6 t^2 /U$, than to the high temperature series result
$T_N^{HT} \simeq 3.83 t^2 /U$.\cite{Hirsch} However, as pointed out by Hirsch
\cite{Hirsch} this could be also an indication that the itinerant nature of
the Hubbard model persists up to large coupling, enhancing the transition
temperature. We cannot settle this debate at large $U$, the TPSC
being inappropriate in this regime, but even at low interaction, our results
are not very close to the Monte Carlo points.

The $U$ dependence of $T_N$ can be traced back using the self-consistency
equation (\ref{autoco}). 
Close to the phase transition the most important
contribution is taken into account by the asymptotic form of the
susceptibility. At $T_N$ $g_{\uparrow\downarrow}=U_{mfc}/U$, and
$\chi_{sp}^{as}$ simply reads
\begin{equation}
\chi_{sp}^{as}(q,0)= \frac 2{\xi_0^2 U_{sp}}\frac 1 {q^2}\,\,\,\,\,\,  ,
\end{equation}
for $q$ measured from ${\bf Q}$. It then follows that the $T_N$ and $U$
dependence of Eq.(\ref{autoco}) can be made explicit
\begin{equation}
 \frac 1 2 \frac {U_{mfc}}{U}= R-T_N \frac 1 {\xi_0^2 U_{mfc} \pi^2} \Lambda
\,\,\,\,   ,
\end{equation}
where $\Lambda$ is the radius of the sphere centered on ${\bf Q}$ where
$\chi_{sp}^{as}$ is defined, and $R$ is a short-hand notation for all terms
which  weakly depend on $T_N$ and $U$. It has been shown\cite{dare} that
at low temperature $\xi_0^2 \propto \frac {\partial^2 \chi_0}{\partial q_x^2}$ 
scales as $T^{-2}$, in the two-dimensional case. This also holds for the 3d
case where 
\begin{equation} \left | \frac {\partial^2 \chi_0}{\partial
q_x^2}\right|_{\bf Q} \simeq \frac{c}{T^2}\,\,\,\,\, ,
\label{xsi0}
\end{equation}
with $c=0.0705$. This behavior extends over a wide temperature
domain including $T_N$.  It then follows that
\begin{equation}
 \frac 1 2 \frac {U_{mfc}}{U} = R - \frac  {T_N^3} {U_{mfc}^2}  \frac
{4 \Lambda}{c \pi^2}  \,\,\,\,\, . 
\label{saturation}
\end{equation}
On the whole range of $T_N$, the scaling behavior of 
$U_{mfc} = 2/ {\chi_0 ({\bf Q}, 0)}$ cannot be neglected and is set, for
$T<t$, by  $\chi_0 ({\bf Q}, 0) \simeq 2 N(0)  \ln(t/T) + \beta$ where $\beta
= 0.38$. Equation (\ref{saturation}) establishes the N\'eel
temperature saturation.

The TPSC formalism has been shown in $2<d<4$ dimensions to be in the same
universality class as the Berlin-Kac spherical model,\cite{dare} i.e. in
the same class as the Heisenberg N-component model with $N \rightarrow
\infty$. The corresponding critical exponents in $d=3$ are:  $\nu=1$,
$\gamma=2$, $\eta=0$, $\alpha=-1$, $\beta=1 /2$, $\delta =5$. Furthermore the
dynamical exponent has been shown to be $z=2$. These exponents  were also
found in the Spin-Hamiltonian approach treated in an Onsager reaction field
theory.\cite{logan} 
The size of the critical region, i.e. for which $\xi \gg 1$, is quite
narrow.\cite{Vilk2} Well above the critical
regime, the behavior of $\xi^{-1}$ with $T$ remains roughly
linear on the whole temperature range studied up to $T \sim 2t$, even for
small $U$, and does not show any crossover to the RPA regime characterized by
$\nu = 1/2$.\cite{Bourbon} The linear behavior of $\xi^{-1}$ at high
temperature can be ascribed to $\xi_0$, (Eqs.(\ref{ksi}), (\ref{ksi02}) and
(\ref{xsi0})).  However for high $T$, the temperature behavior of
$\chi_{sp}^{-1}({\bf Q},0)$ is the mean-field one. Indeed, from Eq.
(\ref{suscepsp}), we have \begin{equation} 
\chi_{sp}^{-1}({\bf Q},0) = \chi_{0}^{-1}({\bf Q},0) - \frac
1 2 U g_{\uparrow \downarrow}\,\,\ . 
\end{equation}
At high temperature $g_{\uparrow \downarrow}$ slowly increases with $T$ (see
next section), and $\chi_{sp}^{-1}$ essentially behaves as $\chi_{0}^{-1}$.

To conclude this section we show in Fig.2 the value of the local
moment $\mu^2 = \left < S_i^z S_i^z \right >$ at $T_N$ as a function of $U$.
TPSC results are compared to DMF results.\cite{Jarrell1}\cite{Jarrell2}
Due to the saturation of $U_{sp}$, the moment does not fully develop
at large $U$ in contrast with the DMF results and with what is expected from
the Heisenberg picture. Nevertheless, 
up to $U \simeq 8$, TPSC gives a very good agreement compared with $d=\infty$
results.

\section{Paramagnetic phase properties}

The central quantity that determines the magnetic properties of the TPSC
approach is the local correlation factor: $g_{\uparrow \downarrow}= \left <
n_\uparrow n_\downarrow \right > / \left < n_\uparrow \right > \left <
n_\downarrow \right >$, which is simply related to the fraction of doubly
occupied sites and to the local magnetic  moment $\left < S_i^2 \right > = 3
(n -\frac 1 2 n^2 g_{\uparrow \downarrow}$). Furthermore,
$g_{\uparrow \downarrow}$ is 
closely related to the self energy 
through the following identity \cite{Vilk2} 
\begin{equation}
\frac 1 2 {\rm Tr} \Sigma G = U  \left < n_\uparrow  n_\downarrow \right > \ \
\ ,  
\label{self}
\end{equation}
where $G$ and $\Sigma$ are respectively the one-particle Green's function and
self-energy. From this exact relation, and in spite of the local nature of
$g_{\uparrow \downarrow}$, one expects a strong dimensional effect due to the 
influence of the magnetic fluctuations on the self-energy $\Sigma$ in
2d.\cite{Vilk2} We are primarily interested in the isotropic 3d case, but we
will also comment upon the 2 and anisotropic 3d cases. 

In Fig.3 the temperature dependence of $g_{\uparrow
\downarrow}$ predicted by TPSC for 3d, is displayed for various values of
$U$. As expected, increasing $U$ is very effective to reduce the fraction of
doubly occupied sites. For $U \gtrsim W$, $U_{sp}$ saturates and thus
$g_{\uparrow \downarrow} \sim 1/ U$ while it should decrease more rapidly with
$U$. In this regime the TPSC approach breaks down. At high temperature
$g_{\uparrow \downarrow}$ slowly increases towards 1, whereas on the low
temperature side, we are limited to $T > T_N$. The arrows locate $T_{min}$,
the minimum of the curves. 

$g_{\uparrow \downarrow}$ has been evaluated in the DMF approach.
\cite{Georges1}, \cite{Georges2} To make quantitative comparisons between TPSC
and DMF results, one needs to use the kinetic energy scale $\sqrt{<
\epsilon_k^2 >}$. Indeed, within the infinite-dimensional model on the
hypercubic lattice, the hopping amplitude is scaled to keep $U / \sqrt{<
\epsilon_k^2 >}$ finite, whereas the bandwidth is infinite.\cite{Metzner}  For
a finite dimension d, averaging over the Brillouin zone leads to $\sqrt{<
\epsilon_k^2 >}= \sqrt{2d}$.  In the comparisons of the coupling dependence of
$g_{\uparrow \downarrow}$ at two temperatures, a quantitative agreement is
found  between the TPSC approach and the DMF results as reported in Fig.4. The
TPSC curves correspond to $T=0.6$ and $T=1.7$, i.e. they surround $T_{min}$,
whereas the infinite-dimensional results extracted from Ref. \cite{Georges1},
where $< \epsilon_k^2 >= 1/2$, are for $T=0.175 \sqrt{12} \simeq 0.6$, and
$T=0.5 \sqrt{12} \simeq 1.73$ respectively. Up to half the bandwidth
($U_{d=\infty} \sim 2$ in $d=\infty$ units), the numerical values are quite
close. 

In both DMF and 3d-TPSC approaches, a
non-monotonous temperature behavior is observed: for $U$ not too strong,
$g_{\uparrow \downarrow}$ first begins to decrease before increasing at higher
$T$. The value of the temperature corresponding to this minimum $T_{min}$
marked by an arrow in Fig.3, is plotted in Fig.5 and compared to the
infinite-dimensional results.\cite{Georges1},\cite{Georges2} They are quite
compatible, since the error bars in Fig.5 for the TPSC results come from the
flatness of the curves. For DMF, $T_{min} \rightarrow 0$ for $U \simeq 4
\sqrt{12} \simeq 14$, whereas in 3d for $U \gtrsim 10$, $g_{\uparrow
\downarrow}$ increases monotonously for $T > T_N$, and cannot be calculated
within our approach for  $T < T_N$. As discussed by
Georges et Krauth, \cite{Georges1} the decrease of $\left < n_\uparrow
n_\downarrow\right >$ reflects an incipient localization tendency when the
temperature increases. This type of behavior is made more acute in the Mott
transition, observed for example for the $V_2O_3$ compound doped with $Cr$:
heating can drive the transition from metal to insulator \cite{McW1} (but the
hypothesis of a structural effect was raised \cite{McKenzie}). The same idea
is at work in the liquid-solid transition in $^3$He, in the so-called
Pomeranchuck effect: the spin entropy contribution is greater in the solid
phase, and increasing the temperature can drive a transition to the solid
state.

From Eq.(\ref{self}) a strong dimensional effect is expected for 
$g_{\uparrow  \downarrow}$, and it becomes interesting to compare the TPSC
predictions for 2, 3 and anisotropic 3d cases to the DMF results. 
To make comparisons between different finite dimensions d, we used again 
the energy scale $\sqrt{< \epsilon_k^2 >}$ rather than naively $W$, (we first used
$W$ and obtained a poor agreement). $\sqrt{< \epsilon_k^2 >}$ is a better
measure of the kinetic energy: it takes into account not only the bandwidth,
but also the particular shape of the density of states. Furthermore, this is
the same scaling as the one previously used for comparisons between $d=3$ and
$d=\infty$. 
In the anisotropic 3d case, the small inter-plane hopping $t_z \ll t$ does not
strongly affect the kinetic energy, and as for the 2d case 
one has $\sqrt{< \epsilon_k^2 >}= 2$.

In Fig.6, we show the temperature dependence of $g_{\uparrow  \downarrow}$ for
the two, isotropic and anisotropic three-dimensional cases in the TPSC
approach. The units on the figure are the 3d ones. 
We first focus on the 2 and isotropic 3d cases. 
The values of $g_{\uparrow  \downarrow}$
cannot be distinguished for 2 and 3d at high temperature. At a
temperature close to the arrows locating the minimum of the isotropic 3d case,
the curves separate. At low temperature in 2d, $U_{sp}$ goes down
when $T$ decreases. This behavior is confirmed by Monte Carlo calculations for
the double occupation factor, \cite{Allen}\cite{Pairault} and is a
manifestation of the incipient antiferromagnetic fluctuations, as can be seen
by inspection of Eq.(\ref{sumSpin}): the increase of the spin susceptibility
integral, leads to a decrease of the double occupation factor. This behavior
is not clearly seen in the 3d case due to the very narrow range of the
critical regime, and to the smaller effect of fluctuations in high dimensions.
The temperature at which the 2 and 3d behaviors separate, close to $T_{min}$,
is higher than the crossover temperature which marks the entering of the
critical regime characterized by a large correlation length,\cite{note} i.e.
$\xi$ does not need to be large to see a strong dimensional effect. 
The non-monotonous temperature behavior of $g_{\uparrow \downarrow}$ observed
in 3d and $d=\infty$ cases, is also present but much less important
in the 2d case for small interaction values, as revealed by a careful
inspection of Fig.6 for $U=2$. In 2d the strong antiferromagnetic
fluctuations prevent a sensible increase of $g_{\uparrow \downarrow}$ when the
temperature decreases. 

In the anisotropic 3d case, a small inter-plane hopping $t_z$ can
stabilize the ordered phase, but has no effect on the value of $g_{\uparrow
\downarrow}$, as shown in Fig.6 for $U=4$, and $t_z=10^{-3}$ (3d units): for
$T > T_N(t_z) \sim 0.18$ the 2d and anisotropic 3d curves cannot be
distinguished.  This behavior is a manifestation of the fact that for these
$U$ and $t_z$ values, the N\'eel transition takes place in the incoherent 3d
one-particle motion regime: the first fermionic Matsubara frequency $\pi T_N$
is much higher than $t_z$. The inter-plane hopping having no
effect on the one-particle Green's function and self-energy, it follows
from Eq. (\ref{self}), that the behavior of $g_{\uparrow  \downarrow}$ is
dominated by two-dimensional effects, even if the critical exponents are the
same for the isotropic and anisotropic 3d models.

In the high temperature regime (typically for $T > T_{min}$), when the
dimension does not have any influence, one can expect that a simple
two-site problem may be instructive. The calculation of $g_{\uparrow
\downarrow}$ in the two-site case is detailed in the appendix, we list here the
main results. The two-site problem solved in the Grand-canonical ensemble,
with a chemical potential fixed to the half-filling value, gives a
good quantitative agreement for the double occupation factor in comparisons
with the TPSC, at high temperature, for low and moderate interaction values.
The two site results also display a non monotonous temperature behavior for
$g_{\uparrow \downarrow}$ with a $U$-dependent minimum.  This minimum is
pushed to high temperature when $U$ increases, in contrast with what was
observed in the TPSC and DMF approaches (Fig.5). However if we freeze the
charge fluctuations which are important in a 2 site problem, working in the
canonical ensemble, the decrease of $T_{min}$ when $U$ increases is recovered. 

From the comparisons between 2d, 3d, $d=\infty$, and the two site
case, it appears that the non-monotonous temperature behavior of $\left < 
n_\uparrow n_\downarrow\right >$ is not a purely dimensional effect, nor a
strong $U$ effect, and that it is overcome by strong antiferromagnetic
fluctuations in $d=2$. 

Finally, we compare our results on finite-size lattices with Monte Carlo
calculations.
In Fig.7, we compare our results for $\mu^2 = \left < S_i^z S_i^z
\right >$ to the Monte Carlo results of Hirsch \cite{Hirsch} on a $4^3$
lattice at $T=0.5$. The TPSC calculations are also done on a $4^3$
lattice. Clearly up to half the bandwidth,
the results are quite compatible. 
In Fig.8 we have plotted the staggered susceptibility as a function of
the site number (4,6,8 and $10^3$), for different temperatures and for $U=6$.
These are compared with the Monte Carlo results of Scalettar {\it et
al.}.\cite{Scalettar}  An overall agreement can be observed except at $T=0.4$
for the intermediate sizes. In the TPSC and in Monte Carlo calculations, a
size effect is clearly seen at this temperature, however for Monte Carlo, the
linear dependence of $\chi_{sp} ({\bf Q},0)$ is the signature of an "ordered"
phase, whereas the transition is not reached yet 
in TPSC. 

\section{Conclusion}

The TPSC approach has been shown in the isotropic 3d case, to give compatible
results in comparison with $d=\infty$ calculations, for the N\'eel temperature
as well as for the local correlation function $\langle n_{\uparrow}
n_{\downarrow} \rangle$. The comparisons with Monte Carlo calculations 
from low up to intermediate coupling regime, except for the
N\'eel temperature which is systematically larger in Monte Carlo, are 
remarkable.

The various results for $\langle n_{\uparrow} n_{\downarrow} \rangle$ obtained
primarily for 3d, but also compared to 2 and anisotropic 3d
Hubbard models shed light on the applicability of the DMF approach. The latter
compares very well with the isotropic 3d case 
for $\langle n_{\uparrow} n_{\downarrow} \rangle$, which is a local quantity
that depends on the dimensionality. That may be seen through Eq.(\ref{self}). 
However, due to the strong ${\bf k}$-dependence of $\Sigma$, the 2d case, or
the anisotropic 3d case for which the N\'eel transition can be reached with a
purely two-dimensional self-energy, are out of
reach of the $d=\infty$ model, as long as the $1 /d$ corrections are not taken
into account in a satisfactory manner. Progress in that direction has been
achieved recently.\cite{vanden}

Finally, let us stress the fact that up to intermediate coupling, the TPSC
gives the same agreement in 2 and 3d, and thus is able to capture the essence
of two very different physical systems. This method is not restricted to
half-filling, nor to the hypercubic case, in contrast with many other
approaches. It then follows that the frustrated antiferromagnetism as well as
the metallic ferromagnetism can be considered. Furthermore a generalization of
this approach can be elaborated in an ordered phase. Work in that direction is
in progress.

It's a pleasure to thank A.-M. S. Tremblay, J.-L. Richard and S. Pairault
for enlightening discussions and key ideas. We are indebted to Yuri Vilk
for providing us with his code enabling the calculations in 2d, and to Steve
Allen for sharing his Monte Carlo calculation results.

\appendix

\section{Two-site results}

In this appendix we detail the calculations of the double occupation factor for
the 2 site problem. From the comparisons between the 2d, isotropic and
anisotropic 3d, and $d=\infty$ results, presented in the main text, and the
observation that, when properly scaled, $\langle n_{\uparrow} n_{\downarrow}
\rangle$ does not depend on the dimension at high $T$, we expect a
good quantitative agreement between the 2 site problem and the infinite lattice
results for any dimensions at high temperature. Apart from a quantitative
agreement, it will be shown that this simple problem roughly explains the 
non-monotonous behavior of  $g_{\uparrow \downarrow}$ as a function of $T$,
and the  $U$-dependence of $T_{min}$ observed in the 3d and $d=\infty$ lattice
case.  

Considering a
two-site problem, with a chemical potential $\mu = U_{ 2 \ sites} / 2$, fixed
to have a mean occupation number per site $n=1$, it is a straightforward
exercise to diagonalize the Hamiltonian, and to evaluate the double occupation
factor. The results for $g_{\uparrow \downarrow}$ for this  problem are
presented in Fig.9 as a function of temperature and various values of the
interaction, along with the 3d TPSC results. 
As previously done for comparisons between different dimensions, interaction
and temperature can be rescaled for quantitative comparisons. With the periodic
boundary conditions for the two sites problem, the kinetic measure is
$\sqrt{\left < \epsilon_k^2 \right >}=2$, as in 2d. For low and moderate
repulsion, and for a temperature typically higher than the 3d $T_{min}$, both
results are close. 
The 2 site 
case not only provides a rough quantitative agreement, but also displays a
non-monotonous behavior as a function of temperature. This one can be easily
understood in terms of eigenvalues $\epsilon$ and eigenstates. In Fig.10 we
have plotted the different values of $\epsilon-\mu N$ as a function of $U_{2
\ sites}$. Table 1 gives a schematic description of the corresponding states
and their contribution to $\left < n_\uparrow n_\downarrow\right >$.

\begin{center}
\begin{tabular}{|c|c|c|c|} \hline
Index& degeneracy& $\epsilon -\mu N$&  $\left < \phi | n_\uparrow
n_\downarrow | \phi \right >$ \\ \hline 
a& 1& $\frac U 2 -\frac 1 2 \sqrt{U^2 +64t^2} - 2 \mu$& $\frac 1 4 - \frac 1 4
\frac {U} {\sqrt{U^2+64 t^2}}$ \\ \hline  
b& 3& $- 2 \mu$& 0  \\ \hline
c& 2& $-2 t +U -3 \mu$& $ 1 / 2$  \\
& 2&  $-2 t - \mu$& 0  \\ \hline 
d& 2& $2 t +U -3 \mu$& $  1 / 2$ \\
& 2&  $2 t - \mu$& 0  \\ \hline
e& 1& $U -2 \mu$& $ 1 / 2$ \\
& 1&  $2U -4\mu$& 1  \\
& 1&  0& 0  \\ \hline
f& 1& $\frac U 2 +\frac 1 2 \sqrt{U^2 +64 t^2} -2 \mu$& $\frac 1 4 + \frac 1 4
\frac {U} {\sqrt{U^2+64 t^2}}$ \\ \hline 
\end{tabular}
\end{center}

TABLE 1. Details of Fig.10. Index refers to the different letters
indexing the curves in Fig.10. For each curve we specify the degeneracy order,
and the value of $\epsilon -\mu N$, the number of electrons being the
coefficient of $\mu$. The last column gives the value of the double occupation
operator in each state. For example the curve c corresponds to 2
three-electron states, with $\left < \phi | n_\uparrow n_\downarrow | \phi
\right > =1/2$ for each, and to 2 one-electron states with no double
occupation. 

For $U_{2 \ sites} < 4$, the first excited state is four-fold
degenerate and corresponds to charge fluctuations (the particle number is
not the same as in the ground state). Two of these four states
correspond to a non-zero double occupation. For higher interaction the
first excited state is three-fold degenerate, corresponding to spin
fluctuations, and has no double occupation. Thus for $U_{2 \ sites} > 4$ the
simple 2 site problem predicts a non-monotonous temperature dependence of
$\left <  n_\uparrow n_\downarrow\right >$. In that case the distance between
the first and second excited sates raises with interaction as can be seen in
Fig.10. It follows that the temperature at which the double occupation factor
begins to increase, grows with the repulsion. This contrasts with the behavior
observed in $d=3$ or $d=\infty$. Nevertheless, the simple 2 sites problem
favors charge fluctuations compared with the thermodynamic limit, and
interestingly the decrease of $T_{min}$ with an increase of $U$, is
recovered if we freeze charge fluctuations taking 2 sites with precisely 2
electrons. If freezing charge fluctuations gives a qualitative agreement for
$T_{min}$ as a function of $U$, the quantitative agreement for $g_{\uparrow
\downarrow}$ in comparisons with the infinite lattice case is not as good as
it was with charge fluctuations allowed.

\begin{figure}
\centerline{\epsfxsize 10cm \epsffile{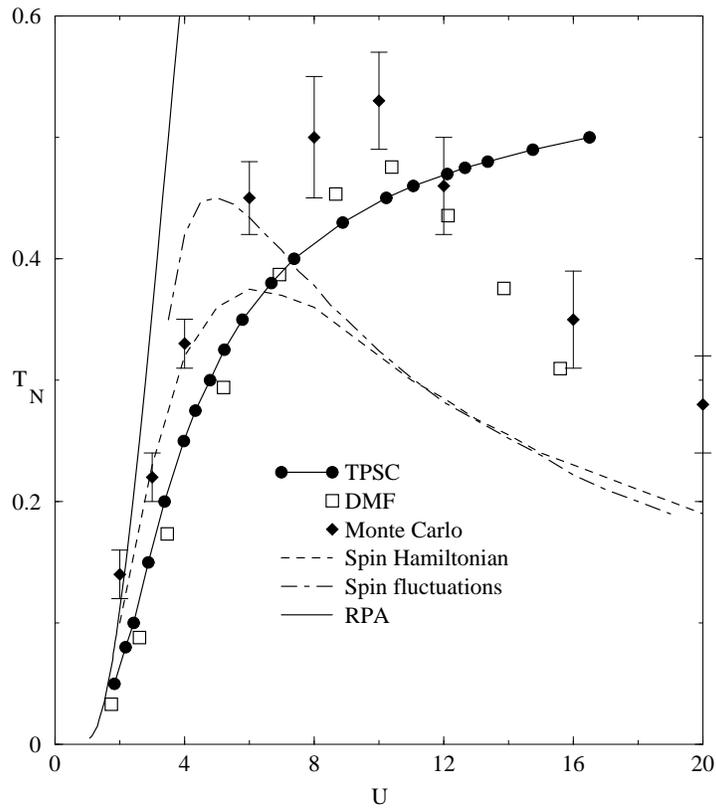}}
\caption{
N\'eel temperature as a function of $U$, for the TPSC formalism compared
with other approaches, see text.}  
\end{figure}

\begin{figure}
\centerline{\epsfxsize 10cm \epsffile{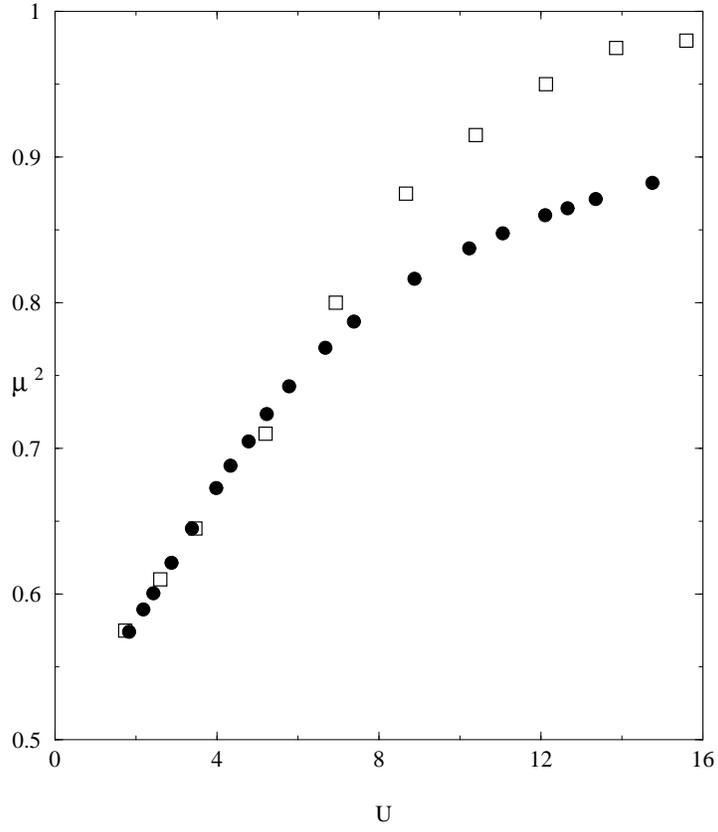}}
\caption{ $\mu^2 = \left < S_i^z S_i^z \right >$ at $T_N$ as a function of
$U$. Black circles correspond to the
$d=3$ case, while open symbols are DMF results, extracted from
Ref. [11]. The energy unit is the 3d one.}  
\end{figure}

\begin{figure}
\centerline{\epsfxsize 10cm \epsffile{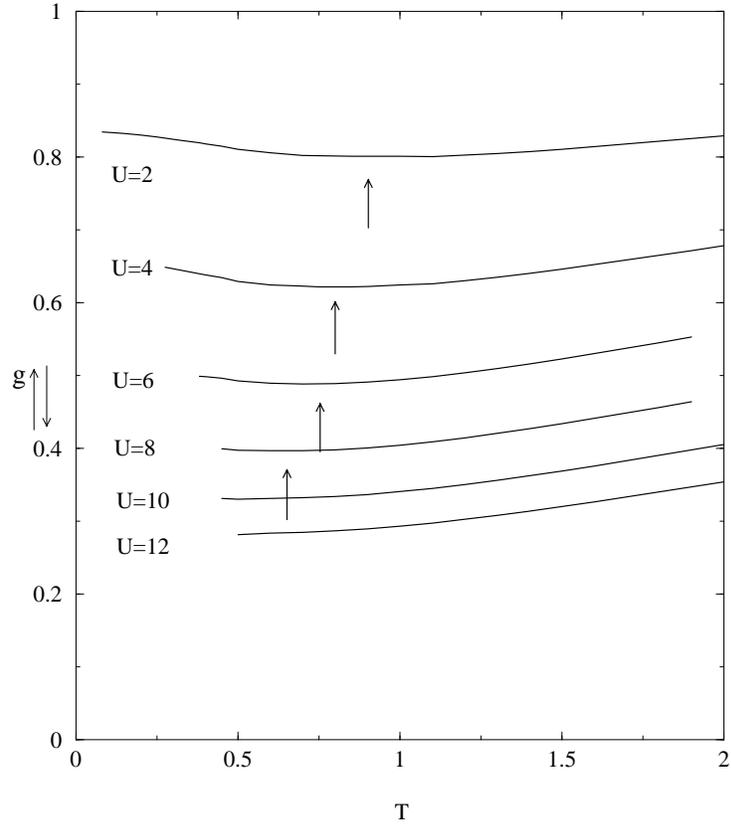}}
\caption{
Double occupation factor $g_{\uparrow \downarrow}= \left < n_\uparrow
n_\downarrow \right > / \left < n_\uparrow \right > \left < n_\downarrow
\right >$ as a function of temperature for different values of the on-site
repulsion. The arrows locate the minimum of the curves.} 
\end{figure}

\begin{figure} 
\centerline{\epsfxsize 10cm \epsffile{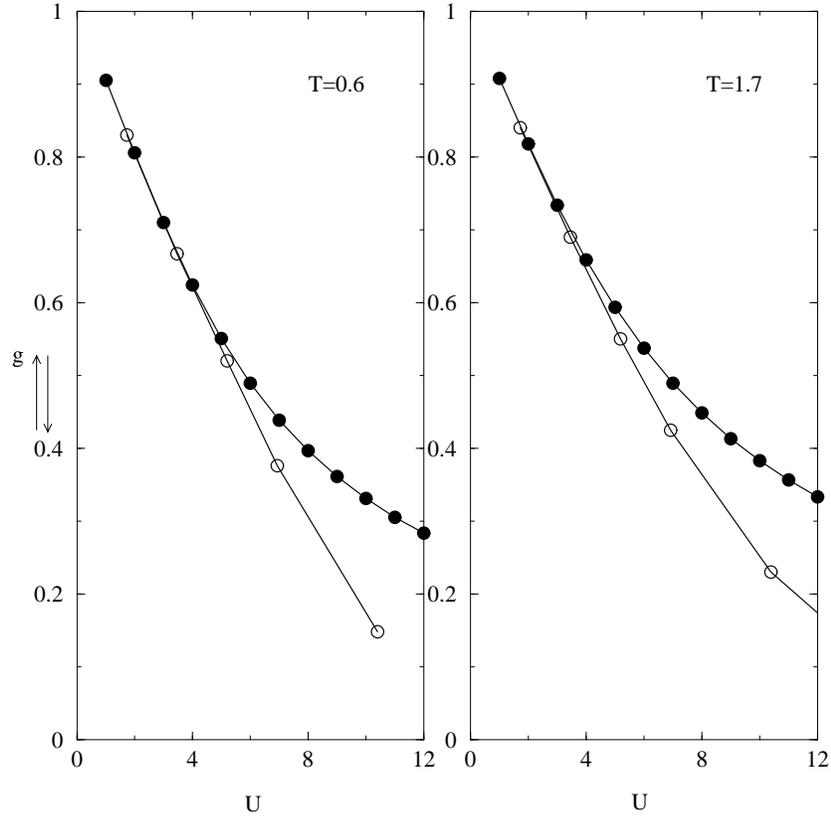}} 
\caption{
Comparisons between 3d-TPSC and DMF for $g_{\uparrow \downarrow}$ as a
function of $U$ for two different temperatures. Black circles correspond to
the $d=3$ case, while open symbols are DMF results, extracted from
Ref.[23]. The energy unit is the 3d one.}   \end{figure}

\begin{figure}
\centerline{\epsfxsize 10cm \epsffile{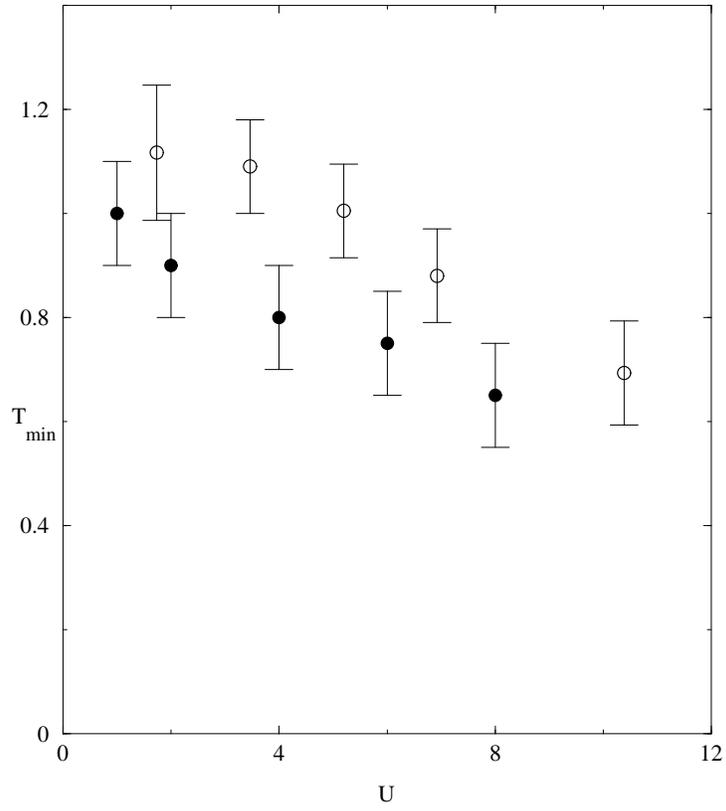}}
\caption{Temperature corresponding to the minimum of $g_{\uparrow
\downarrow}(T)$ for the 3d Hubbard model compared to the DMF results extracted
from Ref.[23]. Black circles are for 3d-TPSC, open symbols for
DMF results. The energy unit is the 3d one.} \end{figure}

\begin{figure} 
\centerline{\epsfxsize 10cm \epsffile{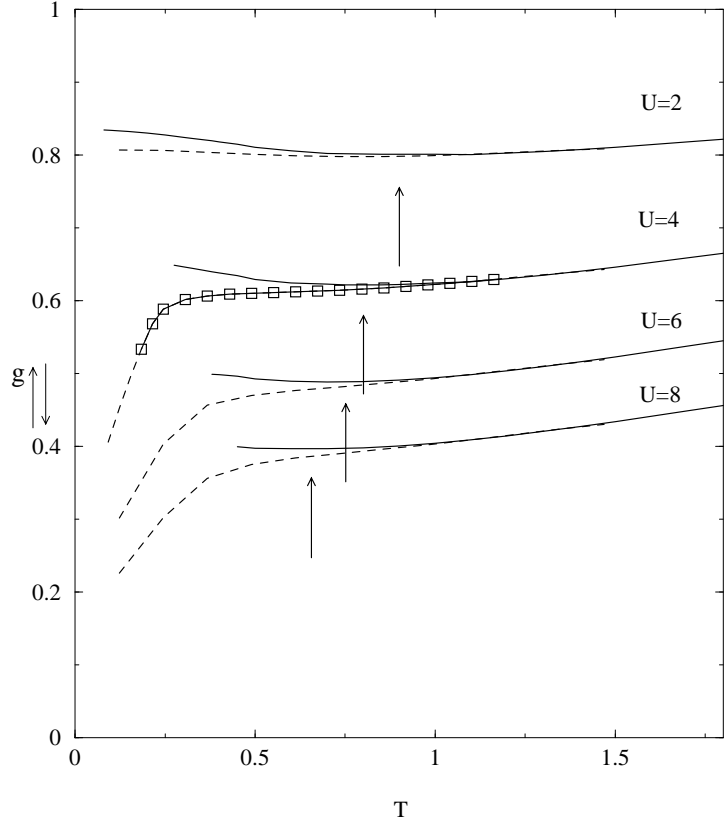}}
\caption{Comparisons between $d=3$, $d=2$, and the anisotropic 3d cases for
$g_{\uparrow \downarrow}$ as a function of $T$, and different $U$ values.
The energy unit is the 3d one (see text). Solid line corresponds to 3d, the
dashed one to 2d. Square symbols are for the anisotropic 3d case, with
$t_z=10^{-3}$ and $U=4$, this curve ends up at $T_N (t_z) \sim 0.18$. The
arrows locate the minimum of the 3d curves.} \end{figure}

\begin{figure}
\centerline{\epsfxsize 10cm \epsffile{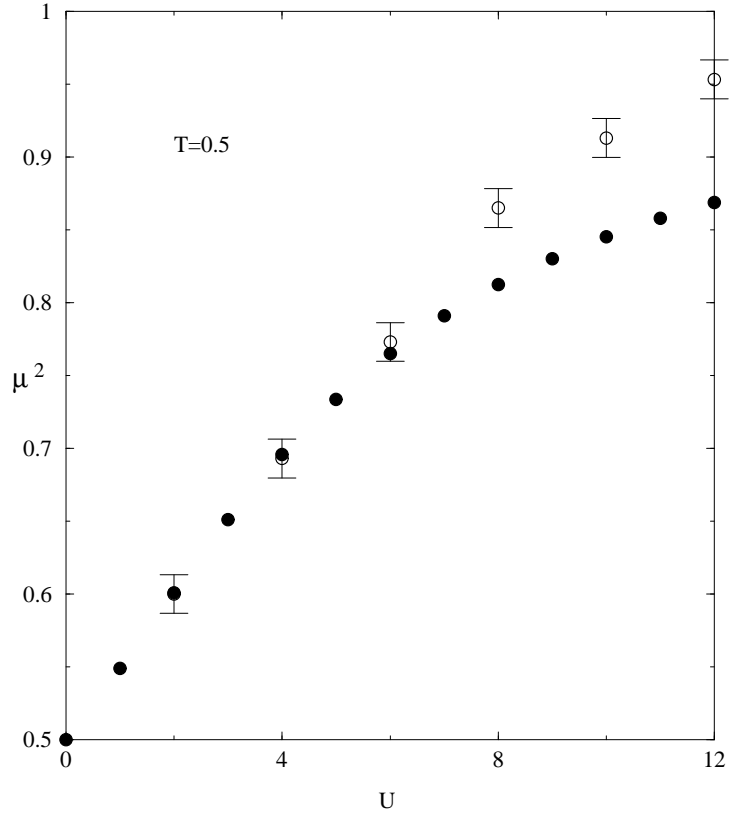}}
\caption{ $\mu^2 = \left < S_i^z S_i^z \right >$ at $T=0.5$ as a function of
$U$. Black circles correspond to the TPSC results on a $4^3$ lattice, open symbols are the
Monte Carlo results on a $4^3$ lattice, extracted from Ref.[20]} 
\end{figure}

\begin{figure}
\centerline{\epsfxsize 10cm \epsffile{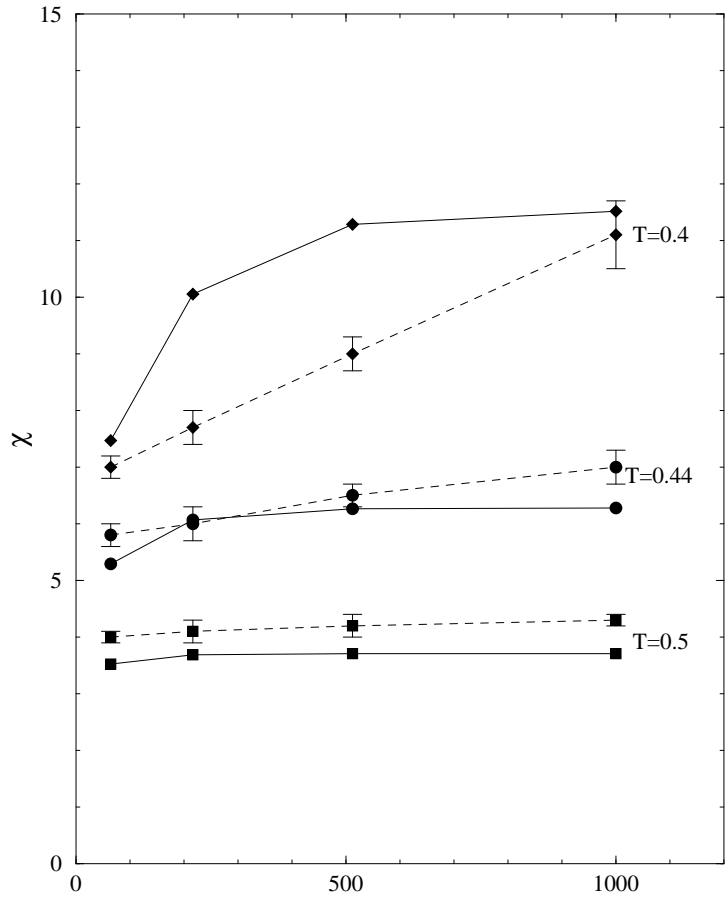}}
\caption{ $\chi_{sp}({\bf Q},0)$ at $U=6$ and different $T$ values, as a
function of the number of lattice sites. Symbols with error bars, linked by
dashed lines are Monte Carlo results from Ref.[10], while symbols linked by
solid lines are the TPSC results.}  
\end{figure}

\begin{figure}
\centerline{\epsfxsize 10cm \epsffile{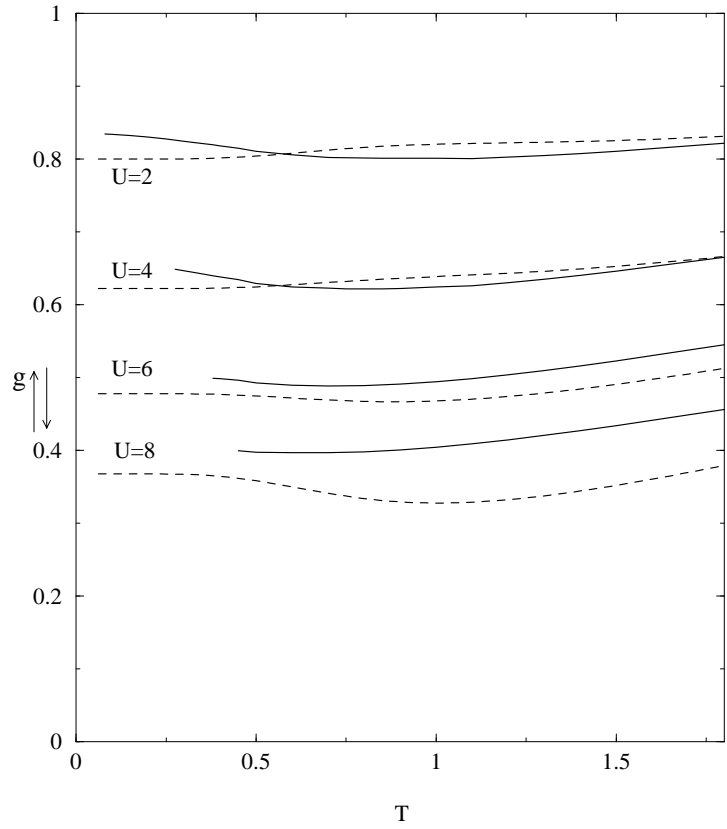}}
\caption{$g_{\uparrow \downarrow}$ as a function of temperature for different
values of the on-site repulsion (solid line) compared to the two-site result
(dashed line).}
\end{figure}

\begin{figure}
\centerline{\epsfxsize 10cm \epsffile{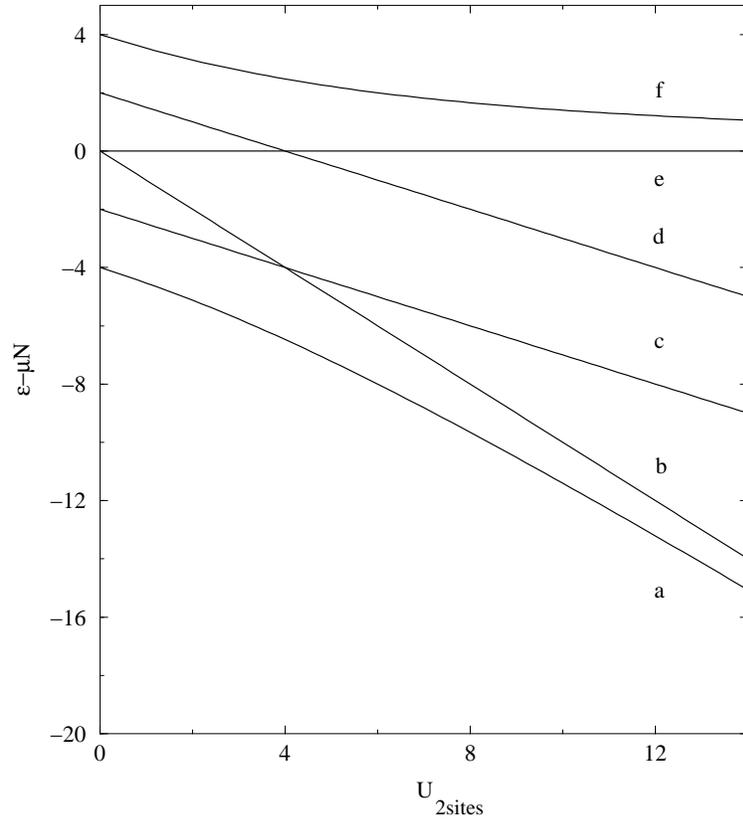}}
\caption{Eigenvalues of the two-site problem $\epsilon-\mu N$ as a function of
the on-site repulsion. Details about these states can be found in Table
1.} 
\end{figure}

\end{document}